\let\oldvec\vec

\documentclass{aa}

\let\vec\oldvec

\pdfoutput=1

\usepackage{graphicx}
\usepackage{amssymb,amsmath}
\usepackage{txfonts}
\usepackage{natbib}
\usepackage{hyperref}
\usepackage{rotating}
\usepackage{color, colortbl}

\renewcommand{\vec}[1]{\mathbf{#1}}

\definecolor{LightCyan}{rgb}{0.88,1,1}

\definecolor{Gray}{gray}{0.9}

\begin{document}

   \title{Coronal type III radio bursts and their X-ray flare and interplanetary type III counterparts
	}

   \author{Hamish A. S. Reid\inst{1,2} and Nicole Vilmer\inst{2}}
   \institute
   {$^{1}$SUPA School of Physics and Astronomy, University of Glasgow, G12 8QQ, United Kingdom \\
   $^{2}$ LESIA, Observatoire de Paris, PSL Research University, CNRS, Sorbonne Universit\'{e}s, UPMC Univ. Paris 06, Univ. Paris Diderot, Sorbonne Paris Cit\'{e}, France }

   \date{}

\abstract
{Type III bursts and hard X-rays are both produced by flare energetic electron beams.  The link between both emissions has been investigated in many previous studies, but no statistical studies have compared both coronal and interplanetary type III bursts with X-ray flares.}
{Using events where the coronal radio emission above 100~MHz is exclusively from type III bursts, we revisited some long-standing questions regarding the relation between type III bursts and X-ray flares:  Do all coronal type III bursts have X-ray counterparts?  What correlation, if any, occurs between radio and X-ray intensities?  What X-ray and radio signatures above 100~MHz occur in connection with interplanetary type III bursts below 14 MHz?}
{We analysed ten years of data from 2002 to 2011 starting with a selection of coronal type III bursts above 100~MHz.  We used X-ray flare information from RHESSI $>6$~keV to make a list of 321 events that have associated type III bursts and X-ray flares, encompassing at least 28\% of the initial sample of type III events.  We then examined the timings, intensities, associated GOES class, and whether there was an associated interplanetary radio signature in both radio and X-rays.}
{For our 321 events with radio and X-ray signatures, the X-ray emission at 6~keV usually lasted much longer than the groups of type III bursts at frequencies $> 100$~MHz.  The selected events were mostly associated with GOES B and C class flares.  A weak correlation was found between the type III radio flux at frequencies below 327 MHz and the X-ray intensity at 25-50~keV, with an absence of events at high X-ray intensity and low type III radio flux.  The weakness of the correlation is related to the coherent emission mechanism of radio type IIIs which can produce high radio fluxes by low density electron beams.  Interplanetary type III bursts ($< 14$~MHz) were observed for 54\% of the events.  The percentage of association increased when events were observed with 25-50~keV X-rays.  A stronger interplanetary association was present when $25-50$~keV RHESSI count rates were above 250~counts/s or radio fluxes of around 170 MHz were large ($> 10^3$~SFU), relating to electron beams with more energetic electrons above 25~keV and events where magnetic flux tubes extend into the high corona.  We also find that whilst on average type III bursts increase in flux with decreasing frequency, the rate of this increase varies from event to event.}
{}


\titlerunning{Simultaneous Type III bursts and X-ray Flares}
\authorrunning{Reid and Vilmer}

   \maketitle

\section{Introduction} \label{sec:intro}

During solar flares, the Sun releases magnetic energy stored in non-potential structures into plasma heating, bulk motions, and particle acceleration. Energetic particles play a key role in the energy release of solar flares since a significant fraction of the released energy is thought to be going into such particles \citep[e.g][]{Emslie:2012aa}.  Energetic particles are also a major driver of solar-terrestrial physics and space weather applications. In particular, radio emission from electron beams produced in association with solar flares provides crucial information on the relationship and connections between energetic electrons in the corona and electrons measured in situ. The most direct and quantitative signature of energetic electrons interacting at the Sun is provided by hard X-ray (HXR) emissions that allow an estimation of electron energy spectra and number density.  Although studied for many years, the connections between HXRs and radio type IIIs are not fully understood.  We statistically investigated the connection between events that show type III bursts in the corona and X-ray flares to further understand their connection.  We also examined the occurrence of the interplanetary counterparts of the `coronal 'type III bursts to explore what electron beam properties and coronal conditions are favourable for continued radio emission in the heliosphere.

Type III radio emissions can be observed over decades of frequency ($1$~GHz to $10$~kHz).  They are characterized by fast frequency drifts (around 100 MHz/s in the metric range) and are believed to be produced by high-energy ($0.05$c-$0.3$c) electron beams streaming through the corona and potentially through the interplanetary space \citep[see e.g.][for reviews]{Suzuki:1985aa,Reid:2014ab}. Type III bursts are one of the most frequent forms of solar system radio emission.  They are used to diagnose electron acceleration during flares and to get information on the magnetic field configuration along which the electron beams propagate.  The bump-in-tail instability produced during the propagation of energetic electrons induces high levels of Langmuir waves in the background plasma \citep{Ginzburg:1958aa}. Non-linear wave-wave interaction then converts some of the energy contained in the Langmuir waves into electromagnetic emission near the local plasma frequency or at its harmonic \citep[e.g.][]{Melrose:1980aa,Li:2014aa,Ratcliffe:2014aa}, producing radio emission that drifts from high to low frequencies as the electrons propagate through the corona and into interplanetary space. In recent years, several numerical simulations have been performed to simulate the radio coronal type III emissions from energetic electrons and investigate the effects of beam and coronal parameters on the emission \citep[e.g.][]{Li:2008ab,Li:2009aa,Li:2011ab,Tsiklauri:2011aa,Li:2014aa,Ratcliffe:2014aa}

Since the discovery of type III bursts and the advent of continuous and regular HXR observations, many studies have analysed the relationship between type III bursts and hard X-ray emissions. The first studies of the temporal correlations between metric (coronal) type III bursts and HXRs above 10 keV were achieved by \citet{Kane:1972aa,Kane:1981aa}. It was found that while about 20\% of the impulsive HXR bursts were correlated with type III radio bursts, only 3\% of the reported type III bursts were associated with HXR emissions above 10 keV. This showed that groups of metric type III bursts were more frequently detected than HXR emissions above 10 keV.  For 70\% of the correlated bursts, it was also shown that the times of X-ray and radio maxima agree within $\pm9$~s. Moreover, the association rate was found to increase when the type III bursts are more intense or when the type III had a larger starting frequency. On the other hand, it was also found that HXR emissions are more often associated with type III bursts when their flux above 20 keV is larger and their spectra are harder.  In a further study based on a larger number of events, \citet{Hamilton:1990aa} confirmed that the association of hard X-ray and type III bursts slightly increases for harder X-ray spectra. They also found that the intensity distribution of hard X-ray bursts associated with type III bursts is significantly different from the distribution of all hard X-ray bursts showing that both kinds of emissions are statistically dependent.  They confirmed on a larger selection of events than \citet{Kane:1981aa} that higher flux radio bursts are more likely to be associated with hard X-ray bursts and vice versa. They also examined whether there is a correlation between the peak count rates of the hard X-ray burst and of the peak flux density at 237 MHz for the associated type III burst. They find no apparent correlation between these two quantities and that there is a large dispersion in the ratio of peak X-ray to radio intensities. 

Other studies have confirmed that type III generating electrons can be part of the same population as the HXR generating electrons.  Indeed, it was shown in different papers that there is a correlation between the characteristics of the HXR emitting electrons (non-thermal spectral index or electron temperature) and the starting frequencies of type III bursts \citep[see][for recent studies]{Benz:1983aa,Raoult:1985aa,Reid:2011aa,Reid:2014aa}. Correlations on sub-second timescales found between HXR pulses and type III radio bursts also strongly support attributing the causal relationship between HXR and radio type III emissions to a common acceleration mechanism \citep[see][]{Kane:1982aa,Dennis:1984aa,Aschwanden:1990aa,Aschwanden:1995aa}.

More statistical studies were performed recently using RHESSI X-ray observations and radio observations from Phoenix-2 in the $100$~MHz to $4$~GHz range. The association between X-ray emissions for flares larger than GOES C5.0 and radio emission (all types of events) was investigated for 201 events \citep{Benz:2005aa,Benz:2007aa}. At the peak phase of hard X-rays,  different types of decimetric emissions (type IIIs but also pulsations, continua and narrowband spikes) were found in a large proportion of the events. For only 17\% of the HXR flares, no coherent emission in the decimetric/metric band was indeed found and all these flares had either radio emission at lower frequencies or were limb events. Classic meter wave type III bursts were found in 33\% of all X-ray flares they were the only radio emission in only 4\% of the events.  The strong association but loose correlation between HXR and radio coherent emission could be explained in the context of multiple reconnection sites that are connected by common magnetic field lines. While reconnection sites at high altitudes could produce the energetic electrons responsible for type III bursts, the reconnection sites at lower altitude could be linked to the production of HXR emitting electrons and some high frequency radio emissions.  

The correlation between hard X-ray and type III emissions can also be examined by combining spatially resolved observations. In some cases, such observations support that type III generating electrons are produced by the same acceleration mechanism as HXR emitting electrons. A close link has been found between the evolution of X-ray and radio sources on a timescale of a few seconds \citep[e.g.][]{Vilmer:2002aa}. However, in other cases the link between HXR emissions and radio decimetric/metric emissions is more complicated. The radio emissions can originate from several locations, one very close to the X-ray emitting site and the other further away from the active region and more linked to the radio burst at lower frequencies \citep[e.g.][]{Vilmer:2003aa}. Such observations are more consistent with the cartoon discussed in \citet{Benz:2005aa}, in which the electrons responsible for type III emissions at low frequencies are produced in a secondary reconnection site at higher altitudes than the main site responsible for X-rays.  In a recent study, \citet{Reid:2014aa} investigated the proportion of events that are consistent with the simple scenario in which the type III generating electrons originate through the same acceleration mechanism than the HXR producing electrons, compared to a more complicated scenario of multiple acceleration regions.  In order to accomplish this, they compared the evolution of the type III starting frequency and the HXR spectral index, signatures of the acceleration region characteristics.  They found that that on a sample of 30 events, 50\% of the events showed a significant anti-correlation.  This was interpreted as evidence that, for these events, there is strong link between type III emitting electrons and hard X-ray emitting electrons.  Such a close relationship was furthermore used to deduce the spatial characteristics of the electron acceleration sites. 

Radio emissions from energetic electron beams propagating in the interplanetary medium have been observed at frequencies below 10 MHz with satellite based experiments since the 1970s \citep[e.g.][]{Fainberg:1972aa,Fainberg:1974aa}.  The most detailed study between coronal (metric) and interplanetary bursts (IP) was performed by \citet{Poquerusse:1996aa} who studied the association between type III groups observed by the ARTEMIS spectrograph on the ground (100-500 MHz) and the URAP radio receiver on the Ulysses spacecraft (1-940 kHz). They found that when there is an association, one single type III burst at low frequencies usually comes from a group of 10 to 100 type III bursts at higher frequencies.  Based on 200 events, they found that 50\% of the events produced both strong coronal and interplanetary type III emissions. They found however that not every coronal type III burst (even if strong) produces an IP type III burst. 

In this paper, we address again, using ten years of data from 2002 to 2011, the long-lasting questions concerning the link between type III emitting electrons in the corona and HXR emitting electrons, and the probability that coronal type III bursts have interplanetary counterparts. Section \ref{sec:eventlist} presents the observations and the methodology used in the study. Section \ref{sec:T3XR} presents results on the link between coronal type III bursts and hard X-ray emissions and Section \ref{sec:T3I3T} discusses the link between coronal and interplanetary type III bursts. Section \ref{sec:discussion} discusses the results in the context of previous but also future observations. Comparisons with predictions of numerical simulations of type III emissions are also tentatively presented.  

\section{Observations and methodology} \label{sec:eventlist}

The present study is based on ten years of observations (2002-2011), from the start of the X-ray observations from RHESSI \citep{Lin:2002aa} to the end of the list of radio bursts reported on observations from the Phoenix 2 spectrometer \citep{Messmer:1999aa} and from the Phoenix 3 radiospectrometer \citep{Benz:2009aa}. Phoenix 2 operated until 2007 and provided observations between 4 GHz and 100 MHz. After 2007, it was replaced by Phoenix 3 which operated between 900 and 200 MHz. The type III events used in our study were extracted from published list of type III bursts\footnote{\url{http://soleil.i4ds.ch/solarradio/data/BurstLists/1998-2010_Benz/}}\footnote{\url{http://soleil.i4ds.ch/solarradio/data/BurstLists/2010-yyyy_Monstein/}}. In both catalogues, only events with `type III flags' were strictly selected. We thus excluded from our list, solar events for which other kinds of radio activity was reported. It should be noted that the period indicated in the catalogue usually contains more than one isolated type III burst and consists typically of several groups of type III bursts that could be separated by quiet periods on timescales of several minutes \citep[][see also the discussion in the next section]{Poquerusse:1996aa}.  In the following, we call this list of type III bursts observed above 100 or 200 MHz (depending on the period) coronal type III bursts. 

Radio fluxes of the type III bursts at different frequencies between 450 and 150 MHz were also computed from the Nan\c{c}ay Radioheliograph (NRH) \citep{Kerdraon:1997aa} observations. The flux was calculated for all events using routines that automatically generated cleaned images of the Sun from the 10 second time integration data.  At each frequency, the location of the maximum radio brightness was searched for and the radio flux was computed on a square window of size 440x440 arcsecs (roughly 7x7 arcmins) around this location. This area is larger than the typical area of a type III burst \citep{Saint-Hilaire:2013aa} at 450~MHz to 150~MHz.  It was chosen because the type III source can move in time and can also include double sources due to the 10 second time integration.  In doing so, the automatically generated flux includes the entire type III sources but does not include significant flux from the quiet sun or other radio sources (e.g. radio noise storms from another active region). 

Radio observations at lower frequencies were also used to search for the counterparts of the coronal type III bursts. Data from the Nan\c{c}ay Decametre Array (NDA) \citep{Lecacheux:2000aa} provided spectra between 80 MHz and 15 MHz. The RAD2 instrument in the WAVES experiment on board the WIND spacecraft \citep{Bougeret:1995aa} provided the observations between 14 MHz and 1 MHz. In the following, we will call the type III bursts observed by WIND below 14 MHz interplanetary type III bursts (IP bursts). 

The list of X-ray flares was obtained from the catalogue of RHESSI X-ray flares. The list is automatically generated and corresponds to an increase of the X-ray count rate in the 6-12~keV range\footnote{\url{http://hesperia.gsfc.nasa.gov/hessidata/dbase/hessi_flare_list.txt}}. Count rates in the 6-12, 12-25, 25-50 and 50-100~keV energy channels were considered in the following analysis. As RHESSI attenuators automatically reduce the X-ray count rate at low energies when it exceeds given thresholds, the more quantitative analysis of the paper is restricted to the use of the 25-50 keV X-ray count rate which is very weakly attenuated even in the strongest attenuator state (A3).

An automatic search for coronal type III bursts and X-ray flares was performed using the Phoenix 2 and Phoenix 3 catalogues for type III flags and the RHESSI X-ray flare list, respectively. The search was restricted to the time window 08:00 UT to 16:00 UT (in order to have NRH observations) and excluded (for the radio observations) periods during which the RHESSI satellite was either in the Earth's shadow (night time) or in the South Atlantic Anomaly (SAA). The automatic search provided a list of 1128 coronal type III bursts and 18,206 X-ray flares above 6 keV, i.e. one order of magnitude more X-ray flares that coronal type III bursts.  The low number of type III bursts with respect to X-ray flares can be explained both by the X-ray energy range considered to build the X-ray catalogue, since the emission in the 6-12 keV range is not necessarily produced by non-thermal electrons and may be attributed to thermal emission, and by the strict selection of type III flags, which excluded type III events associated with other types of radio emissions.

\section{Coronal type III bursts and X-ray flares} \label{sec:T3XR}

\subsection{Do all coronal type III bursts have X-ray counterparts?} \label{sec:T3XR_1}

The automatic comparison of the time intervals of the 1128 selected coronal type III bursts and of the 18,206 X-ray flares showed that for only 581 events, X-ray emission above 6 keV was reported in the RHESSI flare list during the time interval of radio interest.   \emph{The automatic detection of combined X-ray and radio emission thus yields a 52\% association rate between groups of coronal type III bursts above 100 MHz and X-ray emission above 6 keV}.    

\begin{figure*}
  \centering
    \includegraphics[width=0.45\textwidth,trim=20 20 130 120,clip]{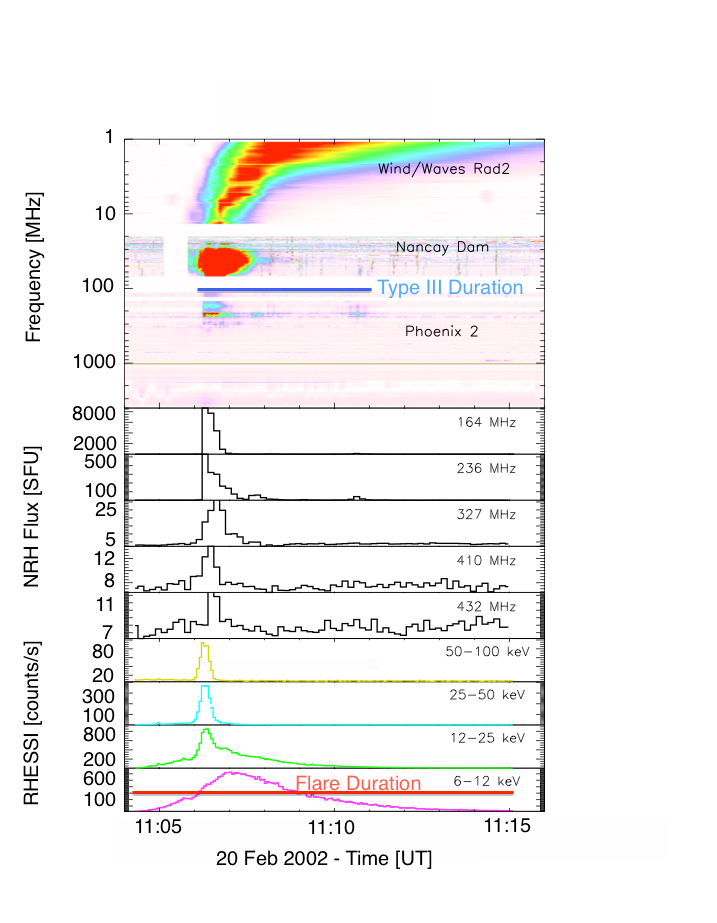}
    \includegraphics[width=0.45\textwidth,trim=20 20 110 100,clip]{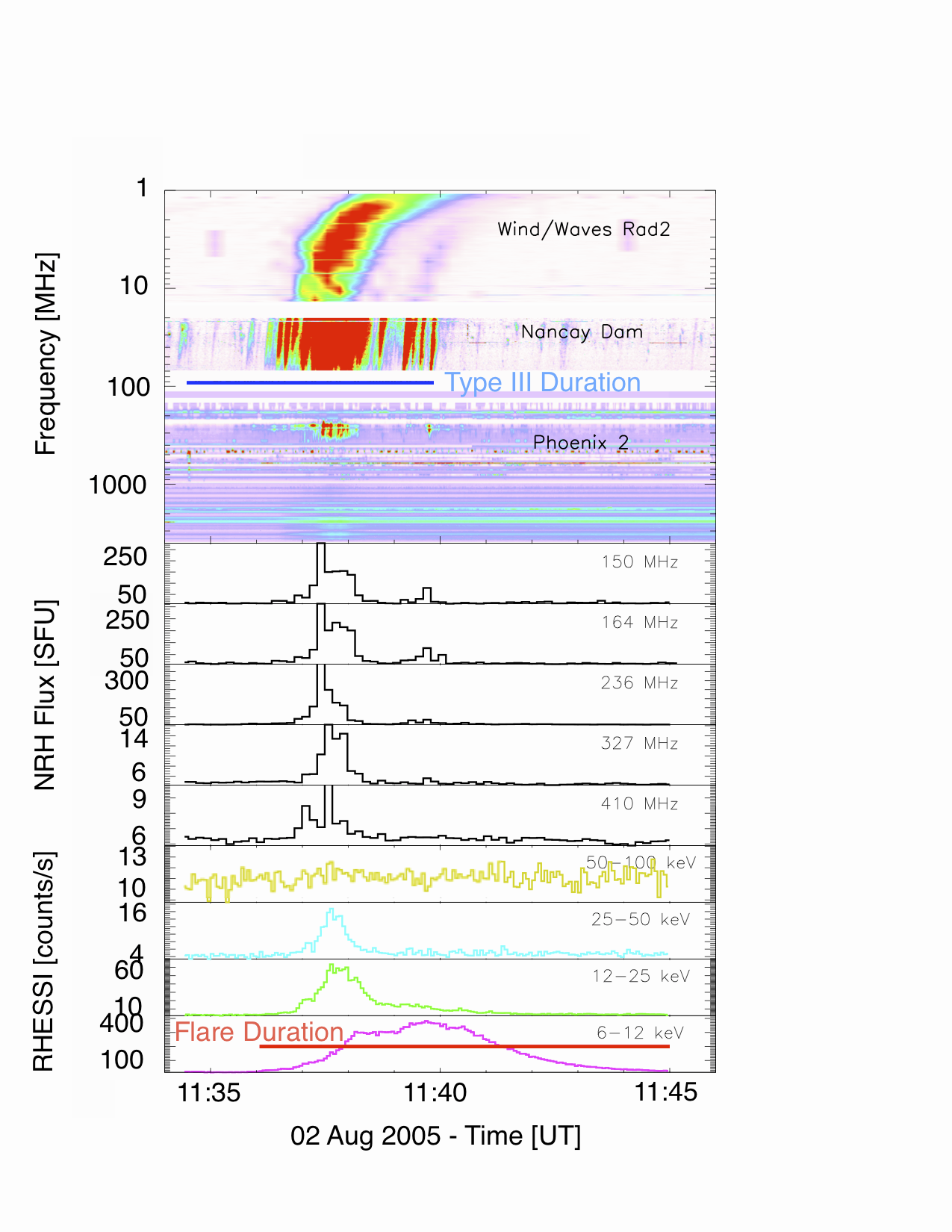}

\caption{Example of two events with associated type III radio and X-ray emissions.  The spectrograms from top to bottom are the WIND/WAVES RAD2 (14 MHz to 1 MHz), Nan\c{c}ay Decametre Array (80 MHz to 15 MHz), and Phoenix 2 (4 GHz to 100 MHz).  The light curves are from the Nan\c{c}ay Radioheliograph (432 MHz to 164 MHz) and RHESSI (6-12 to 50-100 keV).  The blue line represents the duration of the coronal type III bursts as reported in the catalogues.  The red line represents the X-ray flare duration at 6-12 keV as indicated in the catalogue.  There is a linear scale for the light curves.}
\label{fig:example}
\end{figure*}

\begin{figure}
  \centering
    \includegraphics[width=0.45\textwidth,trim=15 20 110 100,clip]{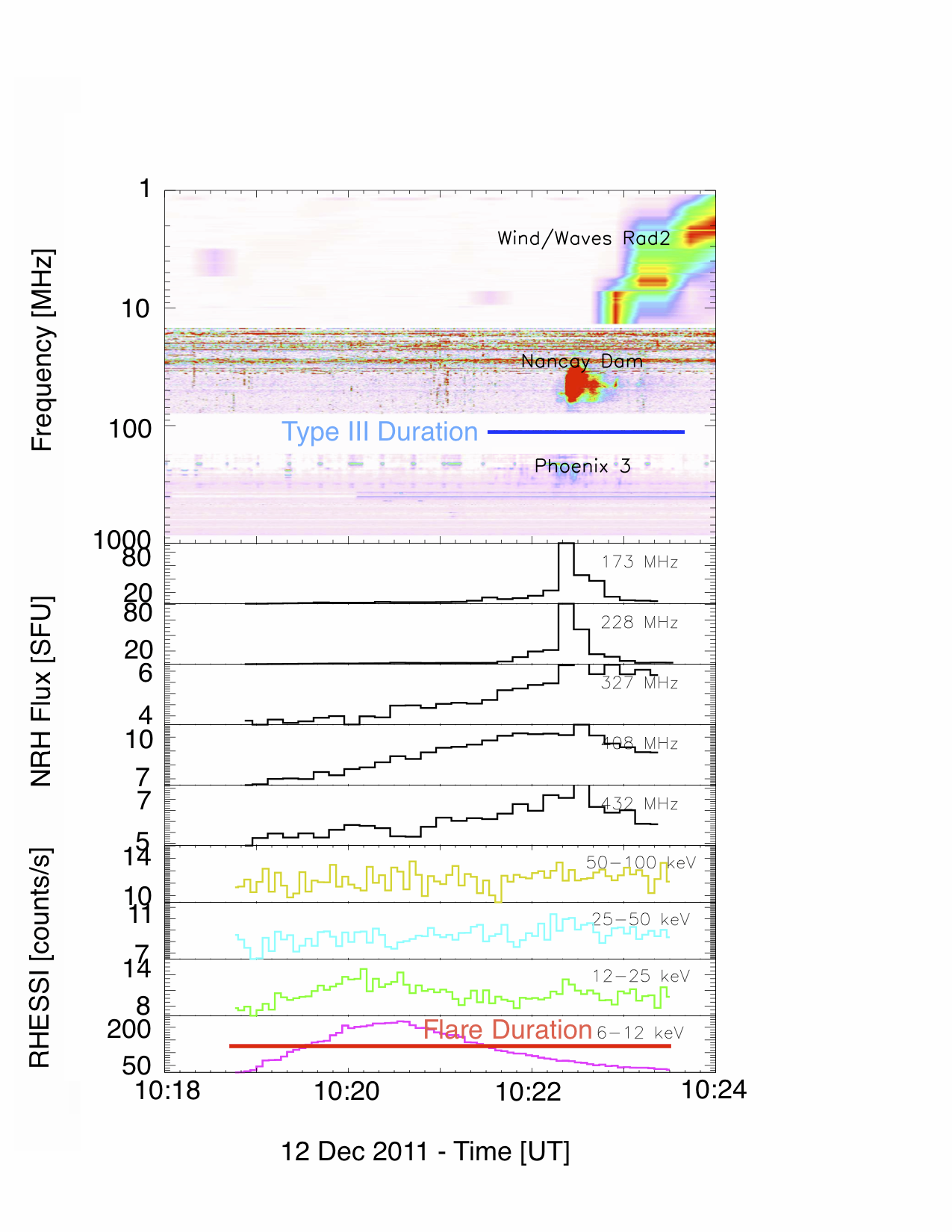}
\caption{Example of an event where the reported coronal type III emission is not associated with the reported X-ray flare.  See caption from Figure \ref{fig:example} for more details on the different plots.  The only exception is that Phoenix 3 was used above 100 MHz instead of Phoenix 2.}

\label{fig:spec_noass}
\end{figure}

Automatic association, while convenient, is fallible. The true association between coronal type III bursts and X-ray flares was thus controlled for all the events by combining the different observations. Figure \ref{fig:example} shows examples of two events for which the association between the reported coronal type III bursts and the RHESSI X-ray flare is excellent.  The aim of the present study is not to search for detailed correlations between the two types of emissions, but just to find global association in time.  Figure \ref{fig:spec_noass} shows an example of an event that has finally been rejected.  The comparison of the relative time profiles of the type III burst and of the X-ray emission shows that the radio burst occurs at the end of the X-ray flare at a time when no X-ray emission is really produced.

Plotting similar figures for other events and checking the true association between coronal type III burst and X-ray emissions significantly reduced the number of associated events. This is because several type III events are reported in the catalogue during one single X-ray flare, bad data occurred in X-rays and/or radio, and, as in Figure \ref{fig:spec_noass}, X-ray flares and type III bursts are not really associated.

A close examination showed that around 100 events had duplicated type III events reported during one X-ray flare and these events were finally merged.  Few events had bad X-ray or radio data.  The most prominent cause of false association was when, upon visual inspection, the X-ray and radio events were not really simultaneous as we show in Figure \ref{fig:spec_noass}.

As a result of the complete check, 321 events were finally found to be associated. \emph{The 321 events gives us a lower bound of 28\% association rate between coronal type III bursts and X-ray flares.}  This is a lower bound because the sample did not include events without reported X-ray flares which could correspond to duplicated reported type III events. For more than half of the 321 associated events, it is found that the reported start and end time of the coronal type III bursts are within the reported duration of the X-ray flare at energies above 6 keV. The general shorter duration of coronal type III bursts can be understood by the fact that type III bursts should be produced by non-thermal electrons whereas X-ray emission above 6 keV is generally predominantly produced by thermal emissions that usually last longer than non-thermal emissions (see e.g. the light curves on Figures \ref{fig:example} and \ref{fig:spec_noass}). Finally, the large majority of the type III-X-ray associated events in our sample are C and B class flares (48\% C and 41\% B). The rest of the events are associated with M class flares (10\%), four X class flares, and 1 A class flare. The very small proportion of large GOES class flares can be understood as an effect of the data selection, since we selected radio events with the type III `flag'. Indeed as recalled in the introduction, GOES flares with classes higher than C5 are almost systematically associated with coherent radio emissions above 100 MHz but in only 4 \% of the events, classical type III bursts are the only radio emission \citep{Benz:2005aa,Benz:2007aa}.

\subsection {What kind of correlation exists between radio and X-ray intensities?}

In this section, the correlation between radio and X-ray intensities is investigated using radio flux measurements from Nan\c{c}ay and RHESSI X-ray count rates in the 25-50 keV range. In the period 2002 to 2011, the NRH observing frequencies underwent some changes as French national protected frequencies changed.  As a consequence, our work is based  on data obtained continuously between 2002 and 2011 in the four frequency ranges defined as  164+173 MHz, 237+228 MHz, 327 MHz, and 408+410 MHz.  The different frequencies within each range are close enough (within 9 MHz) to consider that the physical properties governing the emissions are similar.

Among the 321 events selected in the previous subsection, there were 103 events for which no NRH information was available. For the remaining 218 events, we searched for correlations between the 25-50 keV peak X-ray count rate and the peak radio flux in the different frequency ranges. The peak X-ray count rate and the peak radio fluxes for each event were estimated on a time interval determined by earliest start time and latest end time of the radio and X-ray emissions.  In both X-ray and radio domains, background subtracted peak radio fluxes and X-ray count rates were computed. To automatically determine the background, the median value of each time series was computed as well as the median values of the first and last ten samples. The background was taken as the lowest of these three values.  Such a procedure is especially important for RHESSI observations since the count rates at the start or end of the time series can be strongly affected by the satellite coming out of night time or the SAA.  The background found with this automatic procedure was also checked by manual inspection.  For the few cases when automatic detection of the peak count rates was problematic (e.g. artefacts in the light curves due to changes of attenuators), the time period used to compute the peak count rate was defined manually.  We only kept events for which the peak values were at least 1.5 times the background level in the following analysis.

\begin{figure*}
  \centering
    \includegraphics[width=0.49\textwidth]{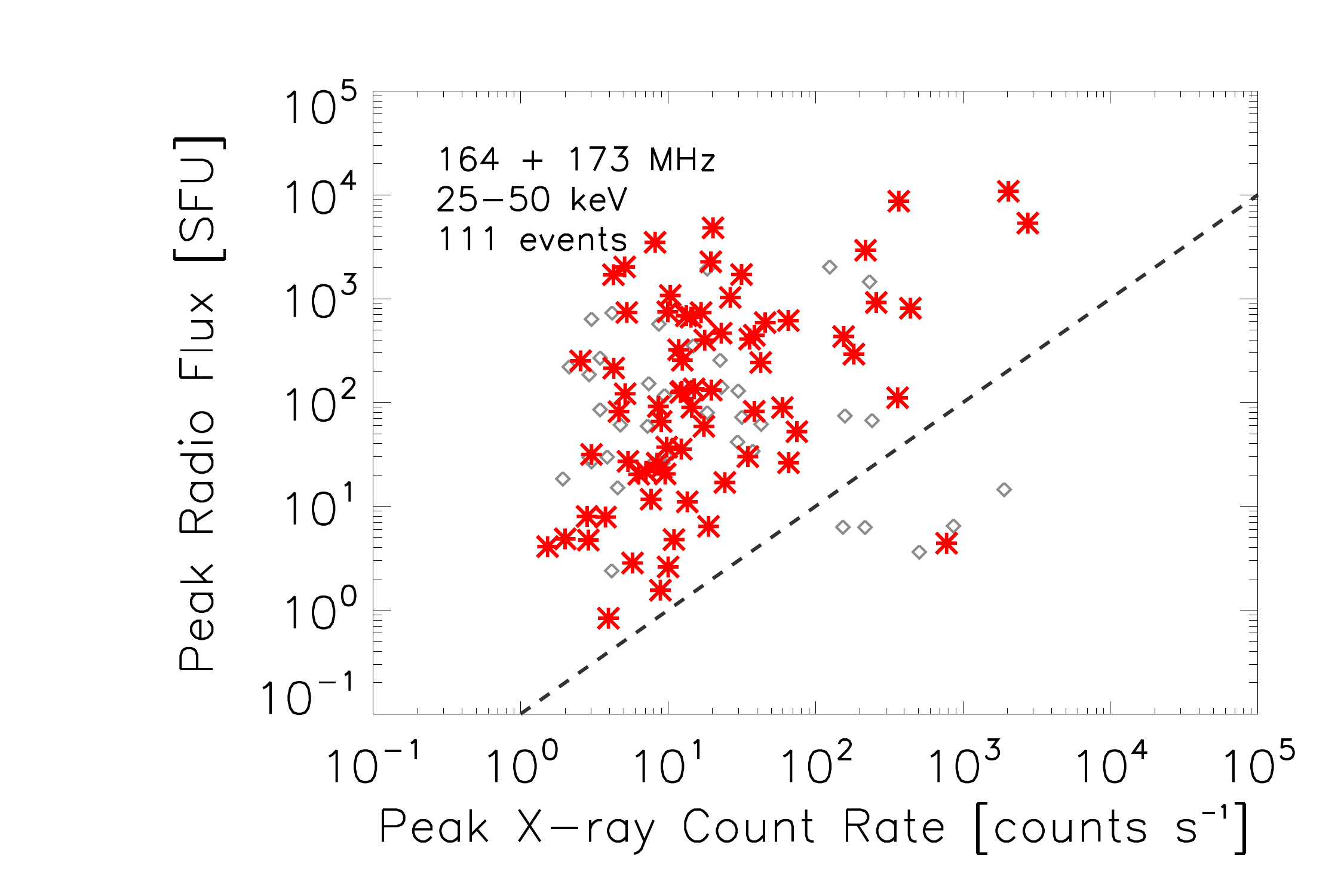}
    \includegraphics[width=0.49\textwidth]{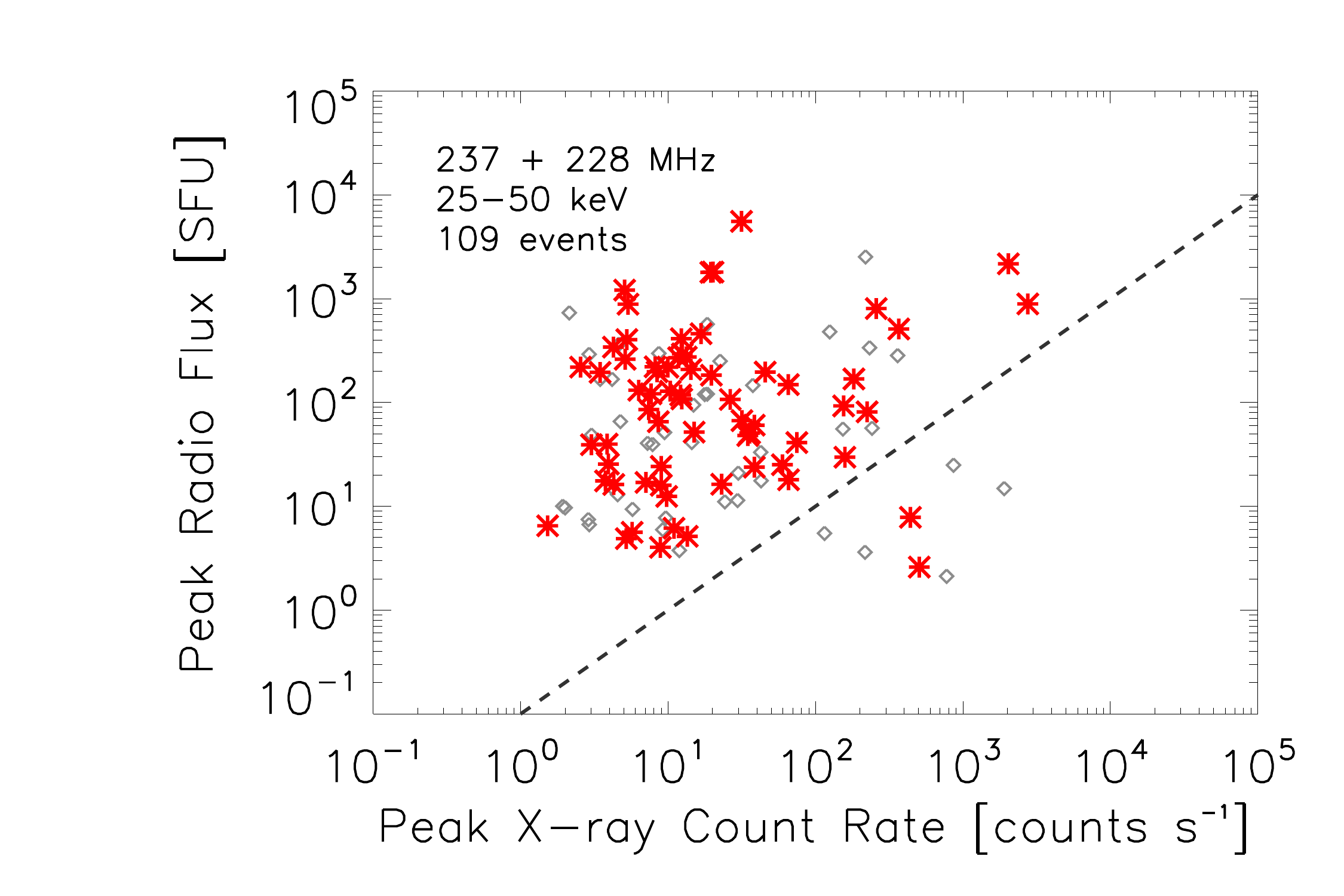}
    \includegraphics[width=0.49\textwidth]{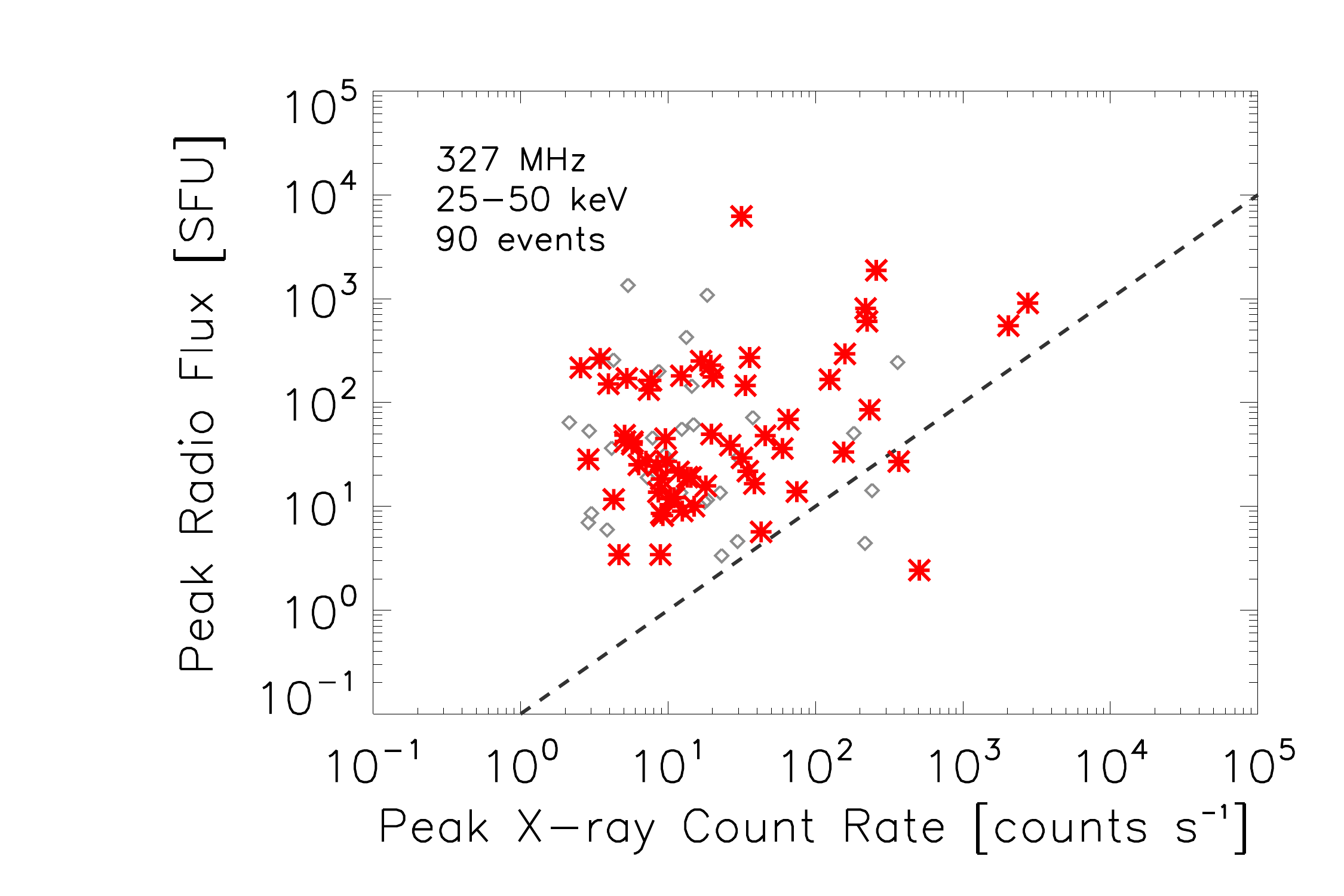}
    \includegraphics[width=0.49\textwidth]{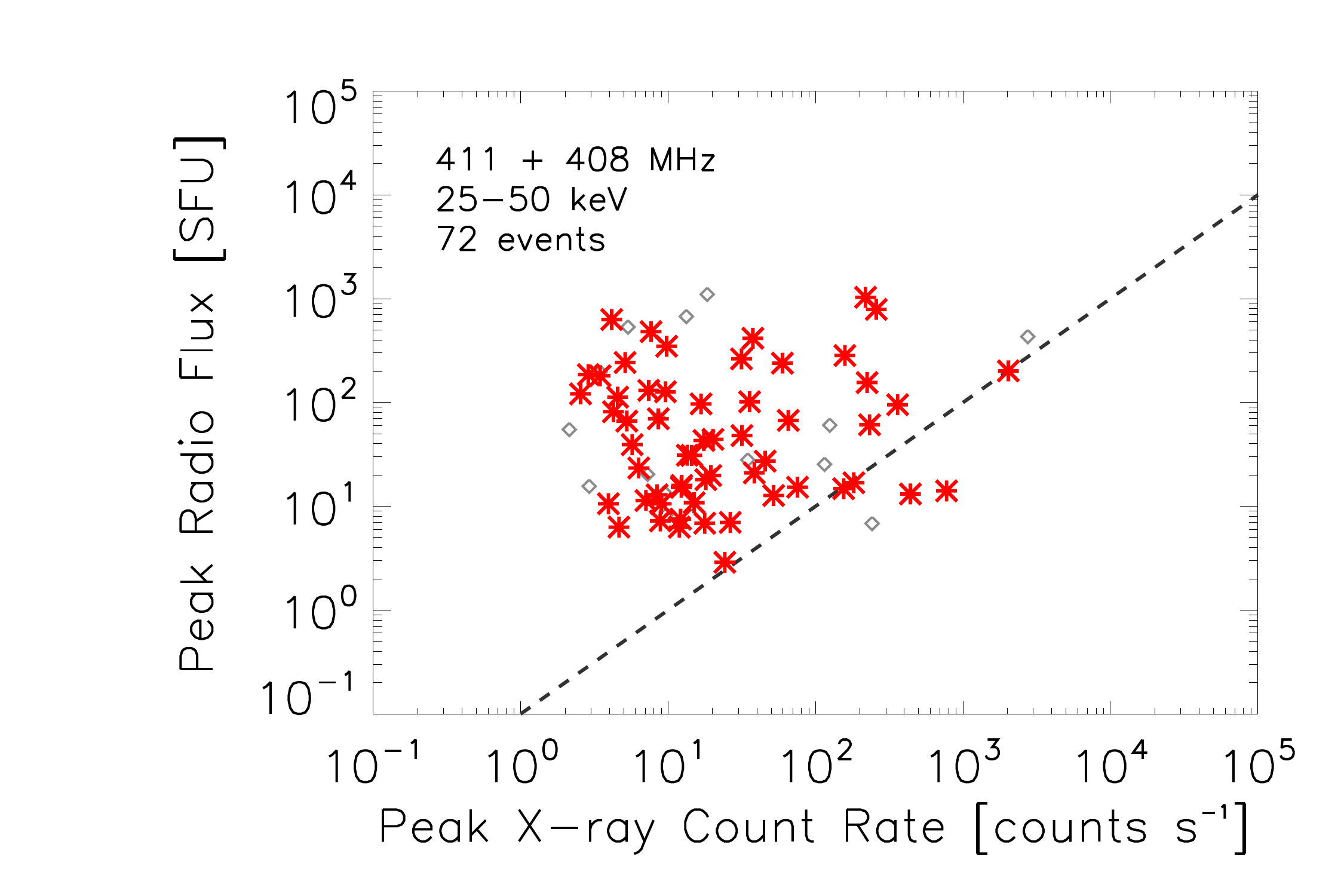}
\caption{Scatter plot of the peak radio flux in the four frequency ranges vs. the 25-50 keV peak X-ray count rate.  The red stars indicate the events where the peak X-ray count rate and peak radio flux are within 40 seconds of each other.  Log-log correlation coefficients of the red stars are 0.09, 0.34, 0.16, 0.46 from around 410 MHz to around 170~MHz respectively.  The dashed line at $y=0.1x$ highlights the general absence of events with high X-ray count rates and low radio flux.}
\label{fig:corr1}
\end{figure*}

Figure \ref{fig:corr1} shows the resulting scatter plot of the peak radio flux in the different frequency ranges versus the 25-50 keV peak X-ray count rate. The red stars correspond to events where peaks in X-ray and radio are within 40 seconds of each other. For these cases, there is an increased confidence that the peaks in X-ray and radio are closely related to each other; as shown on the figure most of the data points correspond to such a situation. The number of events in each frequency range is indicated for each plot. No correlation (correlation coefficient for the log-log values of 0.09 for the red stars) is found between the peak radio flux at 411 and 408 MHz and the peak of the 25-50 keV X-ray count rate. A larger correlation is however observed when comparing 25-50 keV peak count rate with peak radio fluxes at increasingly lower frequencies.  The scatter is still large but the correlation coefficient in the log space is 0.34, 0.16, and 0.46 when we use the subset of events where the peaks must be within 40 seconds of each other (with around 65 events for each frequency range).  The correlation coefficient in the log-space furthermore increases to 0.45, 0.33, and 0.47 only for events with peaks within 15 seconds of each other, but the number of events reduces to around 45.  The most noticeable feature is the clear trend that large X-rays events are associated with large radio fluxes, especially towards the lowest frequencies, i.e. around 230 and 170 MHz, which are indicated by dashed lines in Figure \ref{fig:corr1}.  However, large radio fluxes are not always associated with large X-ray events. This is discussed further in Section \ref{sec:discussion}.

\subsection {Peak radio flux as a function of frequency}

\begin{figure*}
  \centering
    \includegraphics[width=0.51\textwidth]{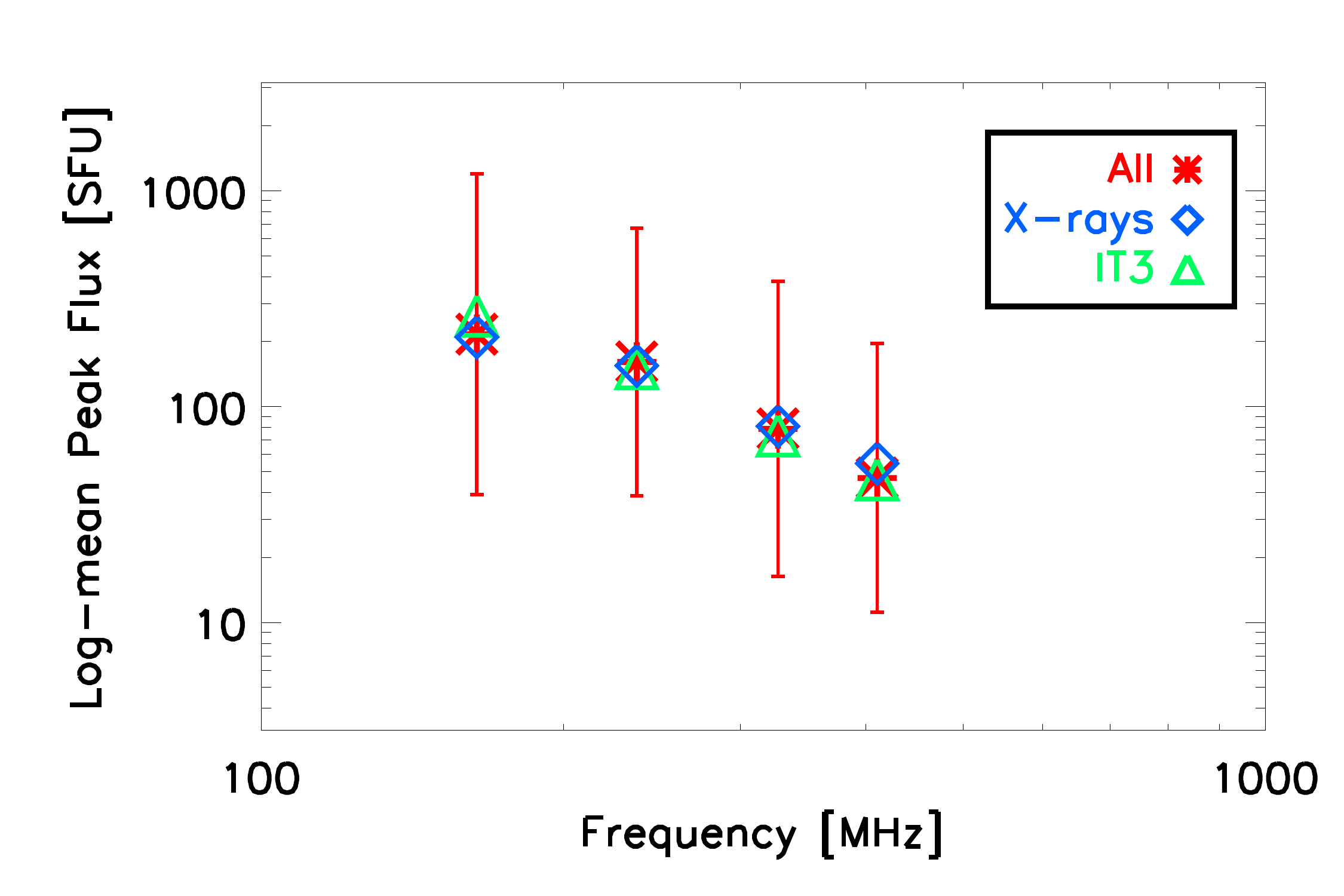}
    \includegraphics[width=0.49\textwidth]{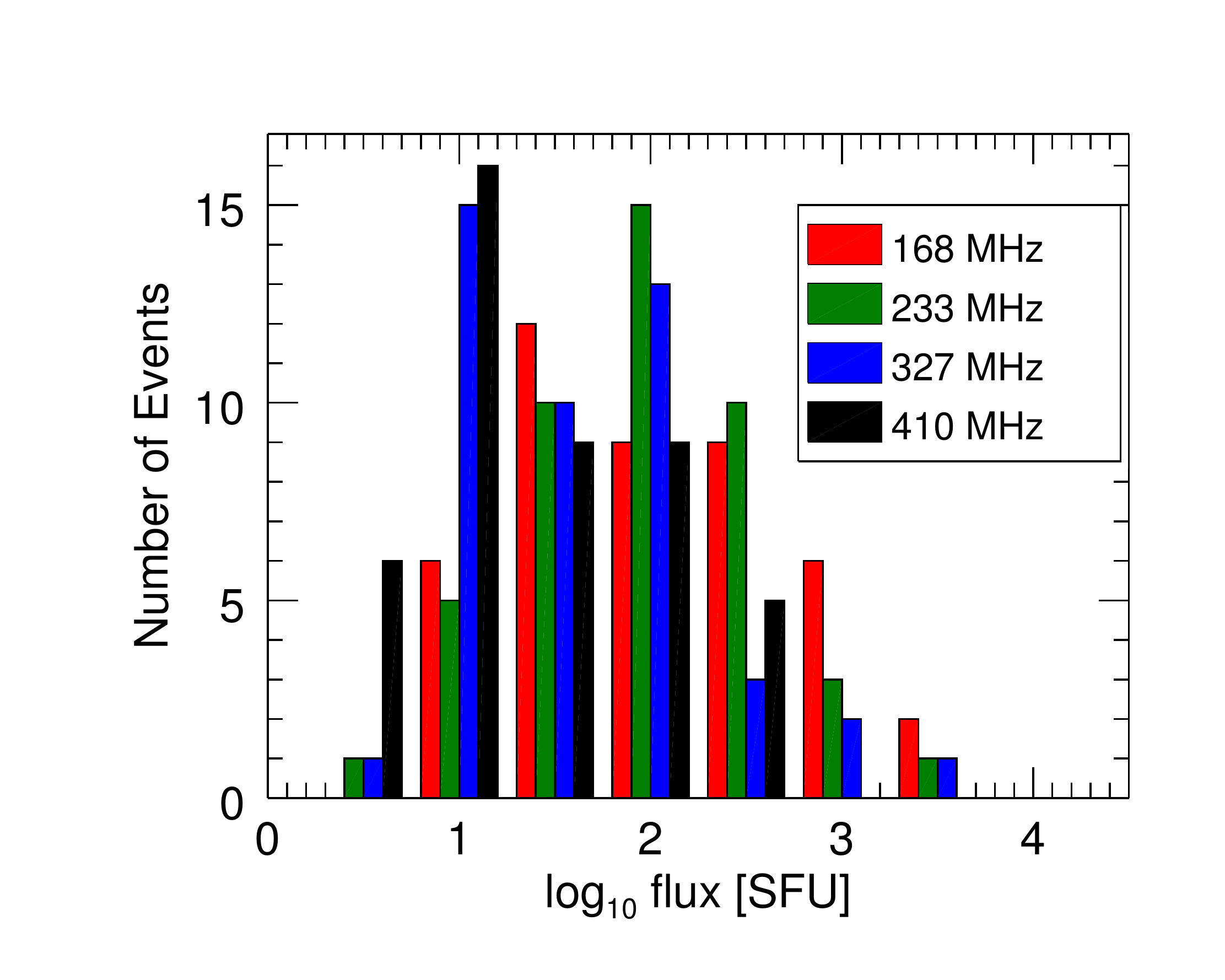}
    \includegraphics[width=0.49\textwidth]{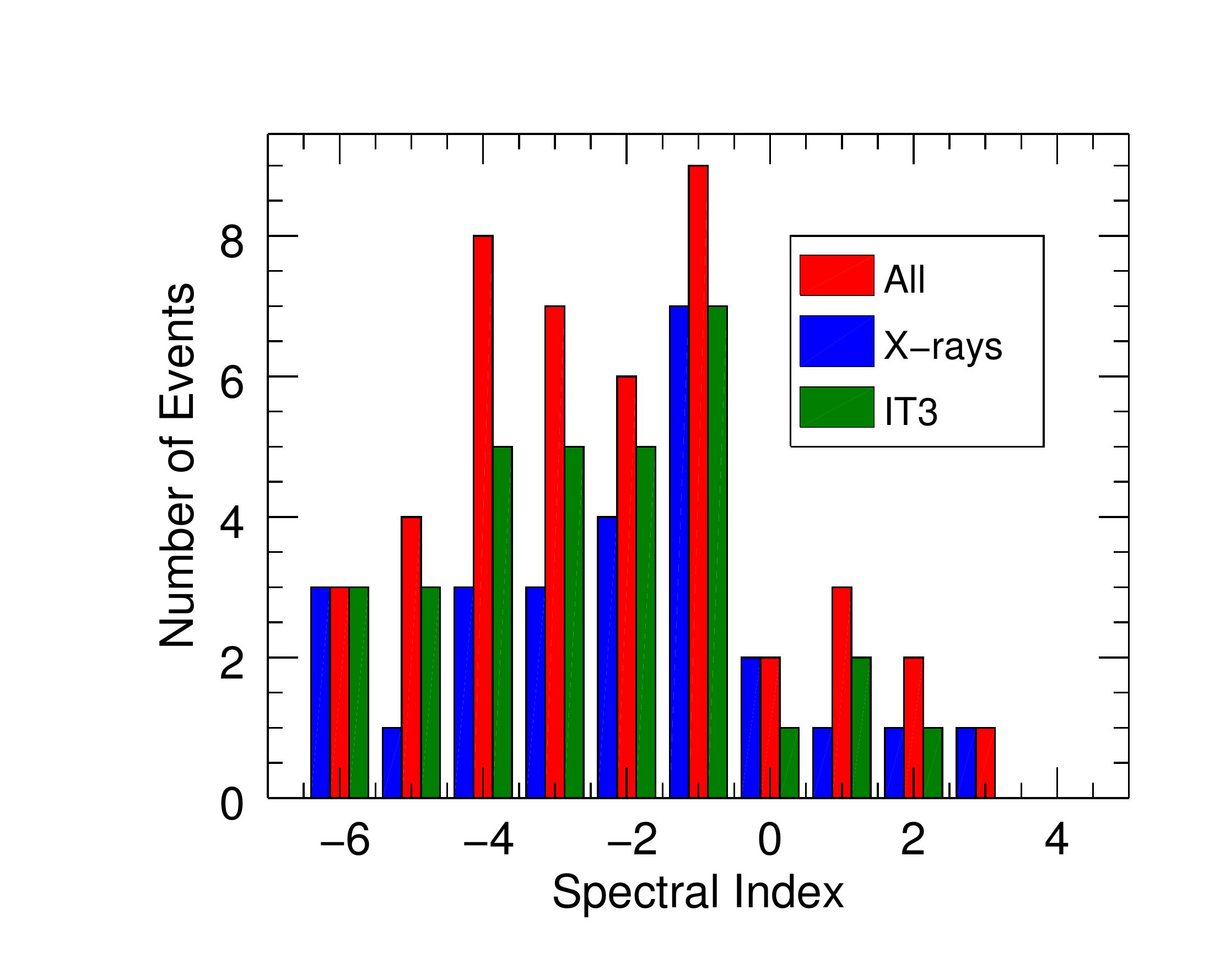}

\caption{Top, a: log-mean peak flux from the 45 (red) type III events that were observed in all four frequency bands from 164 MHz to 410 MHz, where peaks in X-rays or radio are within 40 s of each other.  Errors shown are the standard deviations on the log of the peak fluxes.  Also shown are the subset (32, green) that had interplanetary bursts and the subset (26, blue) that had 25-50 keV X-ray emission.  Bottom left, b: the distribution of the events as a function of peak flux for the 45 type III bursts.  Note the different distributions for each frequency and the large spread in peak flux.  Bottom right, c: the distribution of the spectral indices found by fitting a straight line to the peak fluxes for each event.  All the events (45, red), the subset (32, green) with interplanetary type IIIs, and the subset (26, blue) with 25-50 keV X-rays are shown.}
\label{fig:nrh_peak_flux}
\end{figure*}

Figure \ref{fig:nrh_peak_flux}a represents the evolution with frequency of the log mean radio flux computed on all the associated radio bursts that have signal in all four frequency bands (45 events).  To be considered in the sample, the peak radio flux of the event must be 1.5 times the background in that band.  As in the preceding section, only events where peaks in X-rays or radio are within 40 s of each other are included. Log-means of the radio flux rather than means are used to prevent the strongest but rarest events dominating the results; there exists more than 4 orders of magnitude in the variation of the radio flux.  The error bars in Figure \ref{fig:nrh_peak_flux}a represent the standard deviation of the log fluxes and illustrate the large range of observed flux values.  Figure \ref{fig:nrh_peak_flux}a also shows the same plot for two subsets: events associated with 25-50 keV X-rays (from the points in Figure \ref{fig:corr1}) or events associated with interplanetary type III bursts (IT3).

The log-mean peak radio flux increases as the frequency range decreases. A linear fit using \textbf{mpfitexy} \citep{Markwardt:2009aa} gives a spectral index of -1.78.  However, because of the large scatter in peak flux values, the error on the fit is 2.3.  Figure \ref{fig:nrh_peak_flux}a shows that the log-mean peak flux around 168 MHz is lower than expected from the straight line fit to the other three log-mean peak flux values.  Indeed, the spectral index of the fit is -2.21 if we do not include this frequency band. The spectral index for the events with significant 25-50 keV X-ray emissions is -1.57, while the spectral index for the events associated with interplanetary events is -1.97.

Figure \ref{fig:nrh_peak_flux}b shows the histograms of the peak radio flux for each frequency band.  The form of each distributions is slightly different, with 168 MHz having a negative skewness (skewed to the high flux values) and the higher frequencies having a positive skewness (skewed to the lower flux values).  Moreover, 168 MHz has a higher standard deviation than the higher frequencies.  Both distribution characteristics could be the cause of the slightly lower value for the log-mean peak flux at 168 MHz.

Figure \ref{fig:nrh_peak_flux}c shows the histogram of the spectral indices derived from the evolution of the peak radio flux for all the individual events. The distribution does not look strongly different for events associated with 25-50 keV signal or with interplanetary type III emission below 14 MHz.  Some events have a positive spectral index.  All these results are discussed in more detail in Section \ref{sec:discussion}.

\section{Coronal type III bursts, X-rays, and interplanetary type III bursts} \label{sec:T3I3T}

Interplanetary type III bursts (IT3s) are defined in this study as the extension of the coronal bursts in the frequency range below 14 MHz, observed by the RAD2 instrument on board WIND/WAVES.  Figures similar to Figures \ref{fig:example} and \ref{fig:spec_noass} have been built for all the events to search for significant counterparts of the coronal type III bursts below 14 MHz.  As a result it was found that \emph{for 174 (54\%) of the 321 selected events, a significant signal, i.e. an IP type III burst, was observed below 14 MHz.}

It was furthermore examined whether events with significant X-ray emission above 25 keV had a higher association rate between coronal and interplanetary type III bursts and conversely whether a coronal burst associated with an interplanetary burst had a higher probability of producing significant X-ray emission above 25 keV.  Table \ref{tab:HXRIT3} shows the proportion of events in our sample associated (or not) with IT3 bursts and simultaneously associated (or not) with X-rays above 25 keV. From the different numbers and ratios, it can be seen that if the coronal type III is associated with X-rays above 25 keV, then the association rate with an interplanetary type III bursts is slightly higher (57\%) than for events for which X-rays are only detected below 25 keV.  Conversely, a coronal type III burst associated with an interplanetary type III burst has a slightly higher probability (59\%) of producing significant X-ray emission above 25 keV. These tendencies are discussed in the final section in the context of models relating electron beams and type III bursts.  

\begin{center}
\begin{table*}
\centering
\caption{Proportions of the 321 events that had an interplanetary (IP) type III burst, $25-50$~keV X-rays, both or neither.  The ratios for each row or column are also indicated. }
\begin{tabular}{ c  c  c  c }

\hline\hline
  
321 events & $25-50$~keV X-rays & NO $25-50$~keV X-rays & Ratio\\ \hline

 IP Type III    &  $103~(32\%)$ &  $71~(22\%)$ & $59:41$ \\
NO IP Type III &  $77~(24\%)$  &  $70~(22\%)$  & $52:48$\\
Ratio  &   $57:43$  &  $50:50$ \\
\hline
\end{tabular}
\vspace{20pt}
\label{tab:HXRIT3}
\end{table*}
\end{center}

\begin{figure}
  \centering
    \includegraphics[width=0.99\columnwidth]{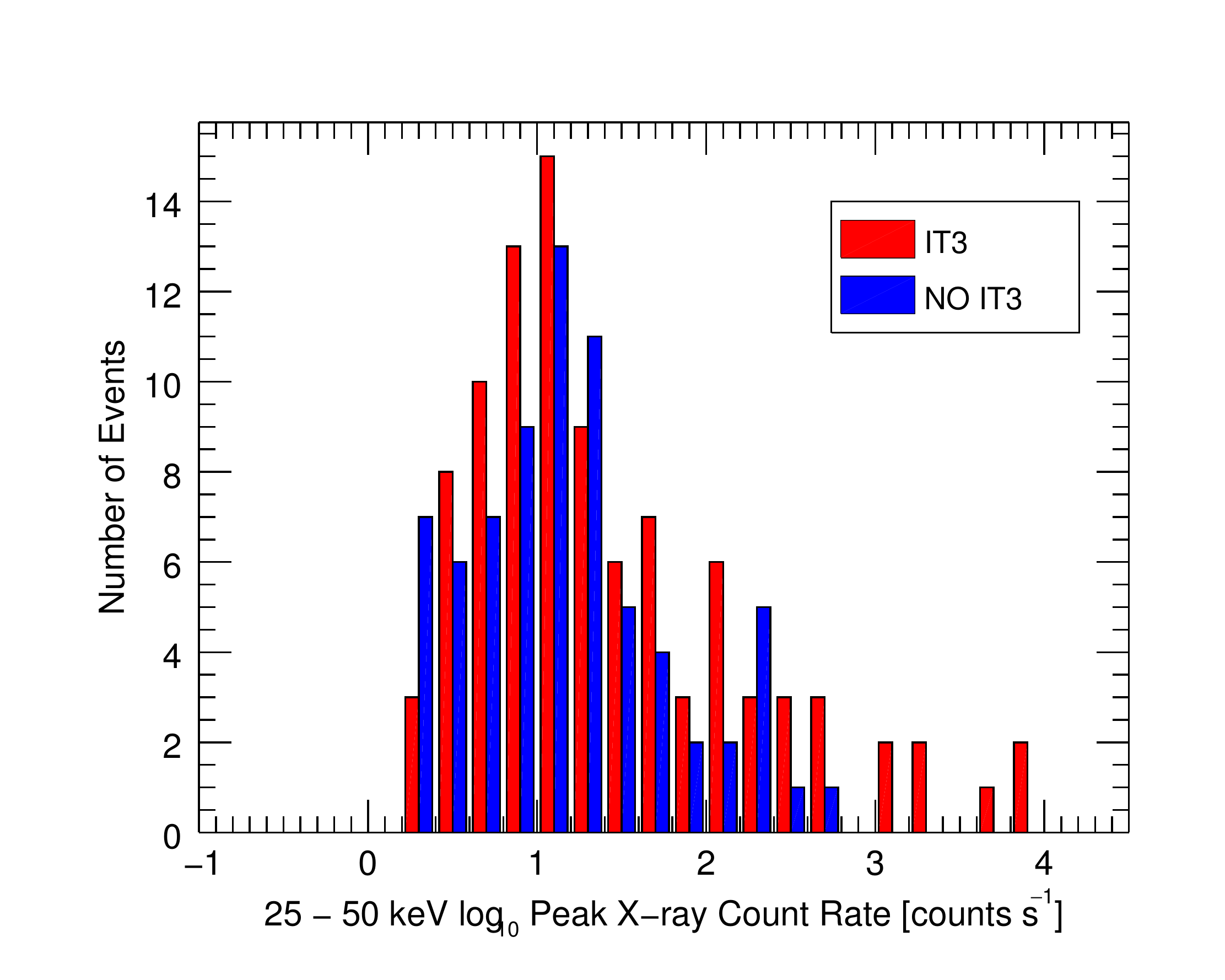}
\caption{Histogram of the background subtracted peak X-ray count rate for the 96 associated events that showed interplanetary type III bursts (IT3s, red fill) and the 73 events that did not show interplanetary bursts (blue line).  The X-ray energy band is 25-50 keV and all events had at least 1.5 times the background count rate.}
\label{fig:xray_hist}
\end{figure}

Exploring the relationship further, Figure \ref{fig:xray_hist} illustrates in more detail the tendency of the coronal type III bursts to be more associated with interplanetary type III bursts (IT3) when they are associated with larger X-ray emissions above 25 keV. The histogram of the background subtracted 25-50 keV peak count rate is represented for events associated or not associated with IT3s.  Only events for which the peak count rate above 25 keV was at least 1.5 times the background were considered. As expected from the above discussion, the distributions are relatively similar, especially at the low count rates.  The log-mean of the two distributions with interplanetary type III bursts (IT3) and without interplanetary type III bursts (no IT3) are 1.44 and 1.17 respectively, corresponding to 27 and 15 counts/s with nearly identical standard deviations. However, above 250 counts/s the distributions are different.  Of the 15 events with high count rates, only 2 do not have an associated IT3 compared to the 13 that do, but the statistics are poor for only 15 events.  In conclusion, the probability of a coronal type III burst to be associated with an interplanetary type III burst depends only weakly on the associated X-ray emission above 25 keV unless the count rate for RHESSI is above 1000 counts/s. In that case, an interplanetary type III burst is likely to accompany the coronal type III burst (but the statistics are still poor).

\begin{figure*}
  \centering
    \includegraphics[width=0.49\textwidth]{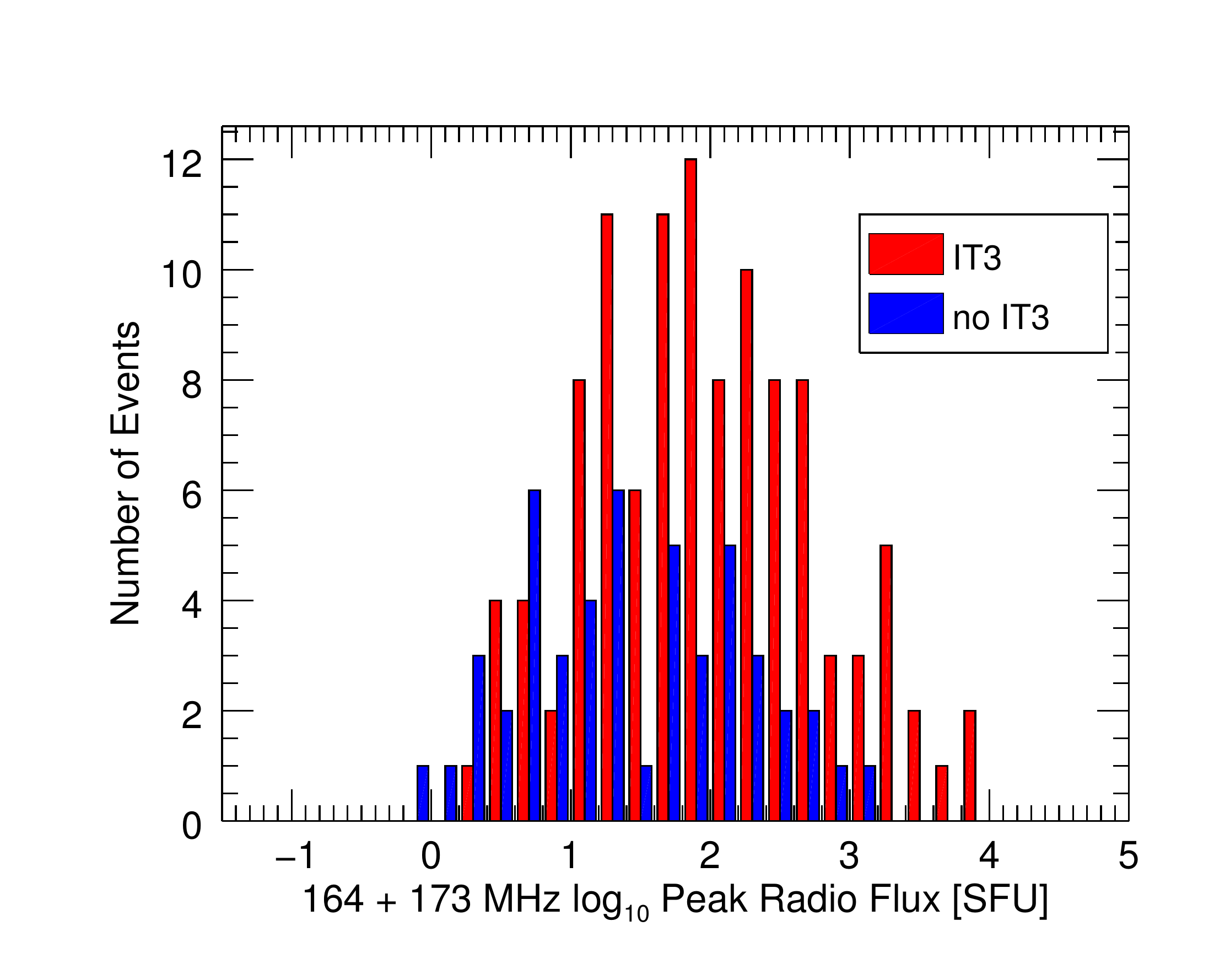}
    \includegraphics[width=0.49\textwidth]{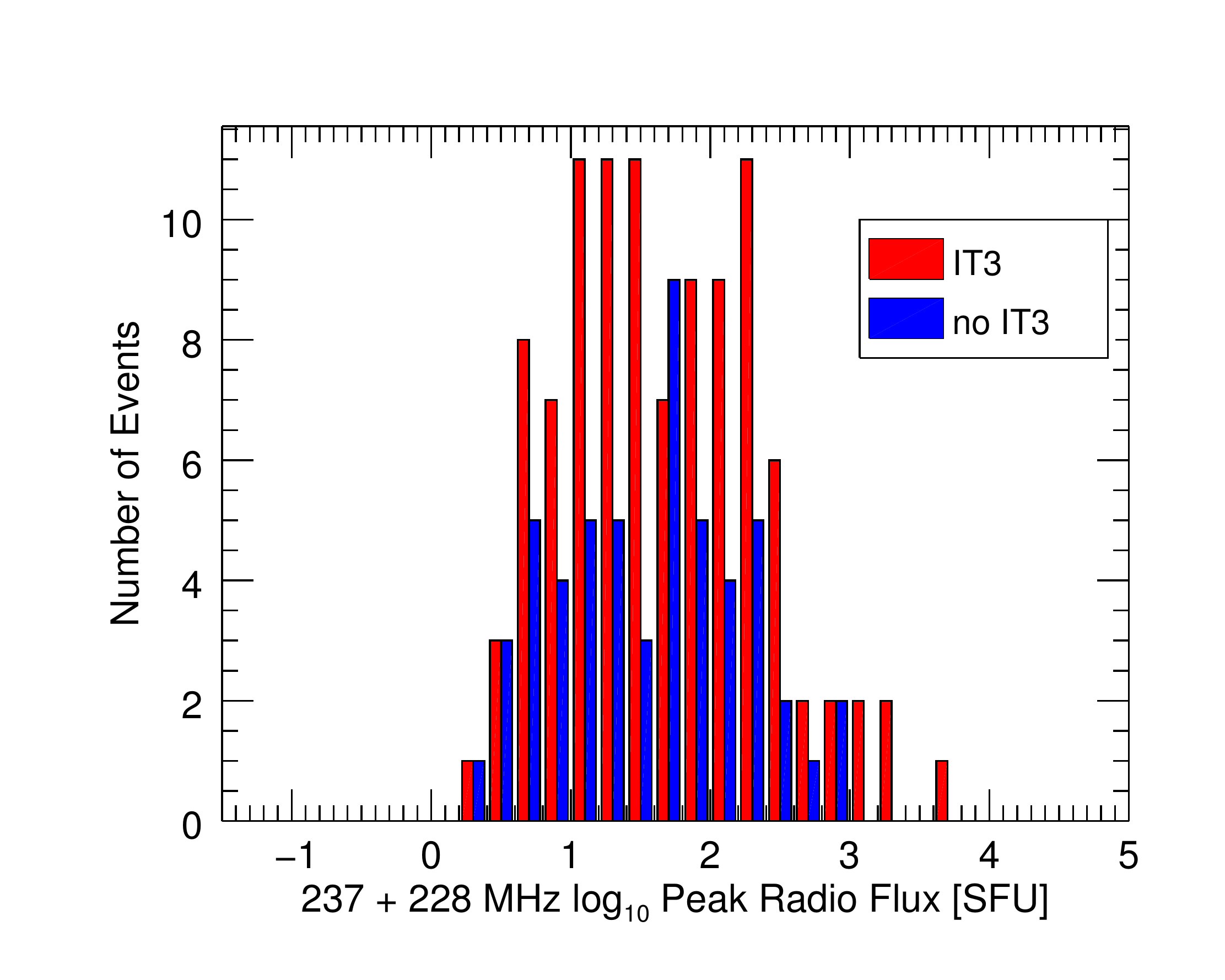}
    \includegraphics[width=0.49\textwidth]{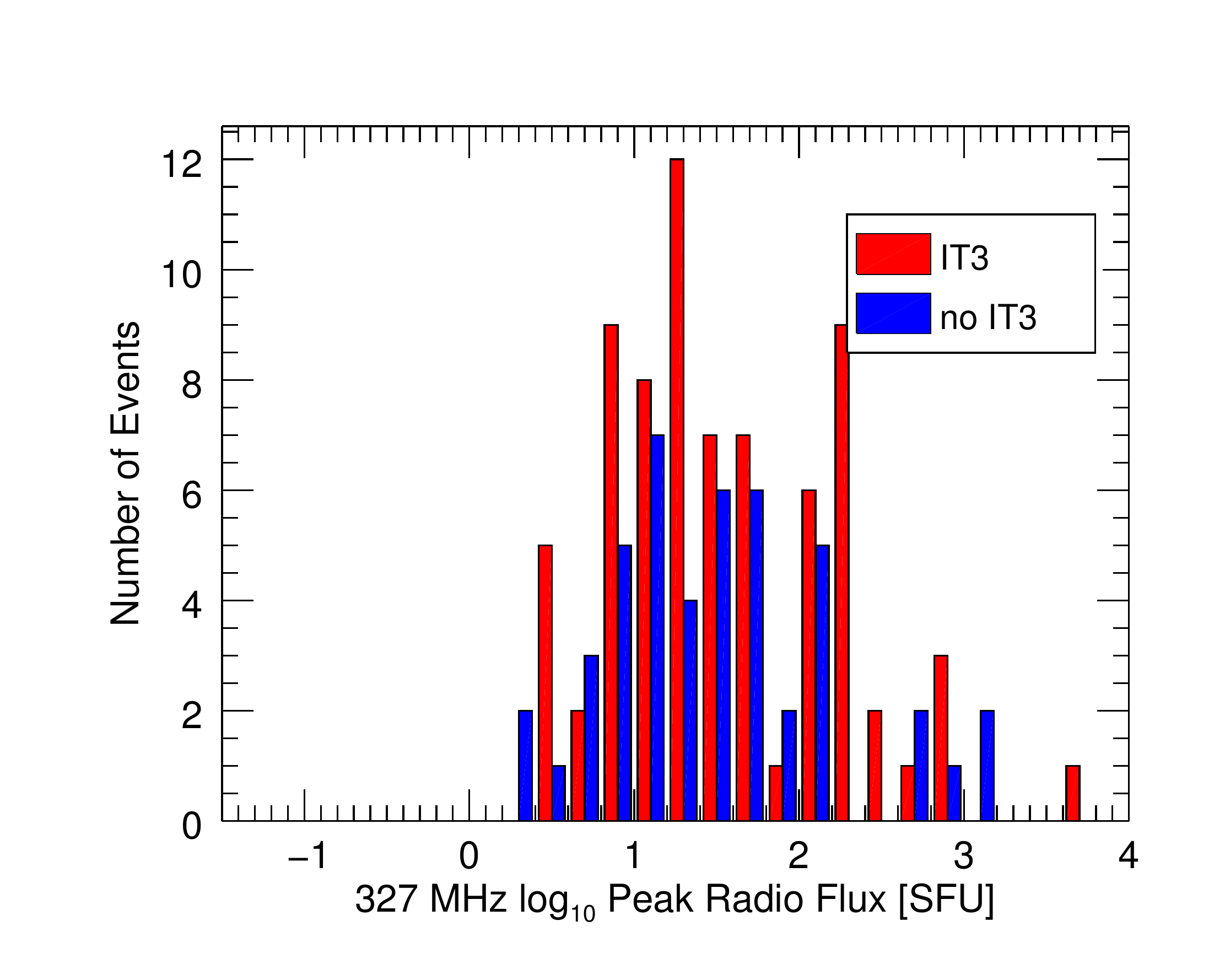}
    \includegraphics[width=0.49\textwidth]{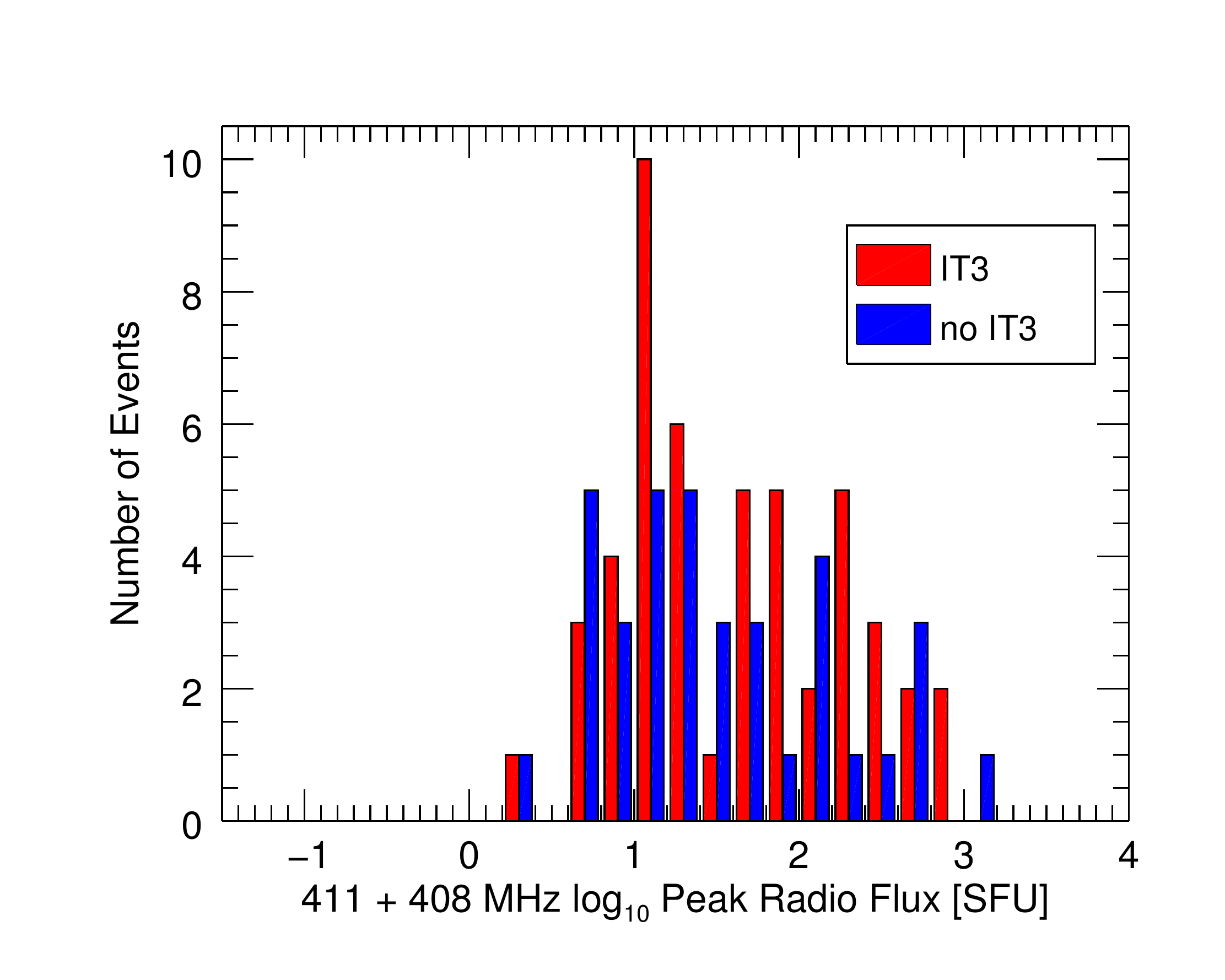}
\caption{Histograms of the background subtracted radio flux for all of the associated events that showed interplanetary type III bursts (red) and did not show interplanetary bursts (blue).  The radio frequencies shown are 164+173 MHz with 109 and 49 events for IT3s and no IT3s, respectively (top left), 237+228 MHz with 103 and 54 events (top right), 327 MHz with 73 and 46 events (bottom left) and 411+408 MHz with 49 and 36 events (bottom right).  All the events had at least 1.5 times the background radio flux.}
\label{fig:radio_hist}
\end{figure*}

Figure \ref{fig:radio_hist} illustrates in more detail the increase of the probability for strong coronal type III bursts to be associated with interplanetary type III bursts.  The histogram of the background subtracted radio flux in the different frequency bands is indeed represented for events associated or not associated with IP type III bursts.  Again only events with fluxes at least 1.5 times the background are plotted.  At the highest frequencies, around 410~MHz the distributions for IT3 or no IT3 are very similar, with similar first and second order moments (log-mean peak flux and associated error, see Table \ref{tab:radio_IT3_noIT3}).  We conclude that the probability of association of a type III burst observed at 410 MHz with an interplanetary type III burst does not depend on its flux at 410 MHz. A similar result is found for frequencies around 327 MHz and 230 MHz, where the log-mean flux is only slightly larger when interplanetary bursts are observed.

At the lowest frequencies around 170~MHz the distributions are however very different (see Figure \ref{fig:radio_hist}).  The log-mean flux is over half an order of magnitude larger when we observe interplanetary type III bursts  than the log-mean flux without interplanetary type III bursts.  The events associated with IP type III bursts have a log-mean of 94 SFU compared to 27 SFU for those not associated with IP type III bursts. We conclude that the probability of detecting an IP type III burst together with a coronal type III burst is strongly enhanced in the case of strong coronal type III bursts around 170 MHz. Although based on a limited number of events, Figure \ref{fig:radio_hist} shows that contrary to what happens at higher frequencies, a high radio flux for type III bursts around 170 MHz is a very good predictor that an interplanetary burst will be observed. This is further discussed in the next section.

\begin{center}
\begin{table*}
\centering
\caption{Log-mean fluxes (in $\log_{10}$~SFU) for the distributions show in Figure \ref{fig:radio_hist} of type III bursts that are and are not interplanetary (have significant emission below 14 MHz).  Values and errors are calculated from the first and second moments of the distributions.  The log-mean is used to avoid biasing the results on the largest events.}
\begin{tabular}{ c  c  c  c  c}

\hline\hline
  
 Log-mean flux  & 164+173 MHz      & 237+228 MHz      & 327 MHz          & 411+408 MHz     \\ \hline
  IT3           & $2.0\pm0.5$ & $1.7\pm0.4$ & $1.6\pm0.5$ & $1.6\pm0.6$ \\
 No IT3         & $1.4\pm0.5$ & $1.5\pm0.6$ & $1.5\pm0.6$ & $1.5\pm0.7$ \\

\hline
\end{tabular}
\vspace{20pt}
\label{tab:radio_IT3_noIT3}
\end{table*}
\end{center}

\section{Discussion and conclusions} \label{sec:discussion}

In this paper, we performed a new statistical analysis on the link between coronal type III bursts and X-ray flares, and on the occurrence of the interplanetary counterparts of coronal type III bursts. The study was based on ten years of observations from 2002 to 2011 using radio spectra in the 4000-100 MHz range for coronal bursts, X-ray data $>6$~keV and RAD2 observations on WIND for the interplanetary type III counterparts.  Based on the RHESSI catalogue of X-ray flares above 6 keV and radio catalogues of pure coronal type III bursts (without other types of radio busts), we investigated the connection between groups of type III bursts and X-ray flares in ten years of data and the link with interplanetary bursts.  We summarize the important results of our study below:
\begin{itemize}
\item The automatic search provided one order of magnitude more X-ray flares at 6 keV than coronal type III bursts. 
\item The large majority of the events in our sample are associated with C and B class flares.
\item Type III bursts above 100 MHz usually start after the X-ray flare is detected at 6 keV and end before the 6 keV emission ceases.
\item A lower bound of 28\% of radio events with only coronal type III bursts had associated X-ray emissions detected with RHESSI above 6 keV.
\item High 25-50 keV X-ray intensities were correlated with high radio peak fluxes below 327 MHz but the opposite was not true.
\item The peak radio flux tends to increase from 450~MHz to 150~MHz but the amount varies significantly from event to event.
\item Interplanetary type III bursts are observed with WIND/WAVES below 14 MHz for 54\% of the coronal type III bursts in our sample. 
\item Events with $>250$~counts/s at 25-50~keV and/or $>1000$~SFU at 170 MHz had a high chance for the coronal type III to become interplanetary. 
\end{itemize}
The finding that one order of magnitude more X-ray flares were detected by RHESSI above 6 keV than pure coronal type III bursts in the same period can be understood from the emission mechanisms.  X-ray emission in the 6-12 keV range is largely produced by electrons with a thermal distribution that is different from the non-thermal electrons that are believed to be the primary cause of type III bursts.  The strict selection of `type III flags' in the catalogue also reduces the number of radio events since it excludes events in which coronal type III bursts would be observed in association with other kinds of coherent radio emissions at decimetric/metric wavelengths.  The thermal origin of the 6 keV X-rays naturally explains the shorter duration of coronal type IIIs as thermal emission is usually observed during the rise and decay of flares in contrast to the non-thermal emission that is localized to the impulsive phase of the flare.

The very small proportion of large GOES class flares in our sample is an effect of the data selection of `pure' type III events. Indeed, as was shown by \citet{Benz:2005aa,Benz:2007aa}, large GOES class flares (i.e. higher than C5) are almost systematically associated with coherent radio emissions above 100~MHz, but classical type III bursts are the only radio signature in only 4\% of these events.

\subsection{Coronal type III and hard X-ray flare connection}

Our results enhance the strong connection already found between electrons that drive coronal type III emission and X-ray flare emission.  We found an association rate of at least 28\% between coronal type III bursts and X-ray flares, which are larger than the 3\% of \citet{Kane:1981aa} and 15\% of \citet{Hamilton:1990aa}.  The larger association rate that we found could be related to our selection of type IIIs as the only radio emission present, considering 6~keV X-rays compared to 10~keV, and instrument sensitivity particularly for X-rays.  The association rate of at least 28\% is further indication that the non-thermal electrons responsible for radio and X-rays are part of a common acceleration process.  However, not all coronal type IIIs have associated X-ray flares.  The lack of simultaneous observations can be due to instrument sensitivity preventing detection of weak simultaneous X-ray emissions associated with type III production or a lack of magnetic connectivity preventing electron transport either up or down in the solar corona from the acceleration site.

The weak log-log correlation between the peak flux of radio emissions at 327-164~MHz and the X-ray count rate at 25-50~keV arises from the notable absence of events with a high X-ray intensity and a low type III radio flux.  This implies that when flares with high X-ray count rates above 25~keV produce coronal type III bursts, the radio fluxes are likely to be high.  However, the correlation is only weak because the converse is not true.  `Coronal' type III bursts with high radio fluxes are also often associated with low X-ray count rates above 25 keV.  This results in a large scatter between X-ray and radio fluxes; this large scatter was already found by \citet{Kane:1981aa,Hamilton:1990aa}.  The large scatter is explained by the increased efficiency of producing a type III radio burst even by a low density electron beam via the amplification of coherent waves.  Conversely the incoherent nature of Bremsstrahlung for producing X-rays is less efficient and the number of high-energy electrons is the primary parameter for producing high rates of X-ray emissions \citep[e.g.][]{Holman:2011aa,Kontar:2011aa}.  The dependency of hard X-rays on the number of high-energy electrons naturally explains the notable absence of events with high X-ray intensity and low type III radio flux.

X-ray flares associated with coronal type IIIs ($>100$~MHz) were found, of which 56\% had $>25$~keV X-rays, which is a much higher ratio than is present for all flares \citep[e.g.][]{Hannah:2011aa}.  A similar result was found by \citet{Kane:1981aa} who showed an increase in the X-ray to type III correlation with the peak flux of X-ray emission.  `Coronal' type III bursts observed in our study need to have a starting frequency well above 100 MHz.  The result reinforces a previously observed property that starting frequencies of type III bursts are linked to the spectral index of the accelerated electron beams \citep{Reid:2011aa,Reid:2013aa,Reid:2014aa}.  The other studies showed that for an electron beam with an injection function of the following form:
\begin{equation}\label{eqn:source}
S(v,r,t) = A v^{-\alpha}\exp\left(-\frac{|r|}{d}\right)\exp\left(-\frac{|t-t_{inj}|}{\tau}\right),
\end{equation}
with an initial characteristic size $d$, injection time $\tau$, and velocity spectral index $\alpha$, the initial height of type III emission depended upon 
\begin{equation}
h_{typeIII}=(d+v_{av}\tau)\alpha +h_{acc},
\end{equation}  
with $v_{av}$ as the average significant velocity of the electron beam and $h_{acc},~(r=0)$ the starting height.  In the case of a harder electron spectrum (lower spectral index), the electron beam becomes unstable to Langmuir wave production after a shorter propagation distance. The starting frequency of the type III emission is thus found to be higher.

`Coronal' type IIIs usually occurred during the impulsive phase of the flare, starting after the 6~keV X-ray rise and ending before the 6~keV X-rays decay.  A similar result was found by \citet{Aschwanden:1985aa} for decimetric emission between 300-1000~MHz, many of which included type III bursts.  Additionally, the peak flux in emission between X-rays and type IIIs is usually closely related in time.  For the events of our sample for which $>25$~keV emission was detected, around two-thirds of the events have peak fluxes within 40 seconds, decreasing to around half of the events for peak fluxes within 15 seconds, which is just above the limit of time cadence used in our analysis.

\subsection{Peak radio flux verses frequency}

We investigated the evolution in frequency of the type III log-mean peak radio flux.  Although there is a very large dispersion in the spectral index found (see Figure \ref{fig:nrh_peak_flux}c), the peak radio flux is found as a mean (in log-space) to increase with decreasing frequency with a mean slope of -1.78.  The evolution of the log-mean radio flux with frequency furthermore depends on the association of the event with either significant hard X-ray emission above 25 keV (slope of -1.57) or with an interplanetary burst (slope of -1.97).  The large scatter we observed shows that the frequency evolution of type III bursts is varied from event to event.  The slopes are significantly different from the slope of -2.9 found on a survey of 10,000 type III bursts observed with the Nan\c{c}ay Radioheliograph \citep{Saint-Hilaire:2013aa}.  The discrepancy could be related to the fact that our events only consider bursts that emit in all NRH frequencies as opposed to a statistical average over all bursts.

Our results can be taken as further evidence that, in general, type III bursts are more numerous and stronger at low frequencies than at high frequencies.  Physical reasons for the increase include the onset and increase of the bump-in-tail instability with velocity dispersion, the lower background plasma density reducing collisional damping and increasing the Langmuir wave growth rate, and the lower frequency radio waves escaping the corona more easily.  The increase in radio flux with decreasing frequency is consistent with what was observed by \citet{Dulk:1998aa} who found that the spectrum of a type III burst in the $3-50$~MHz range has a negative spectral index.  More recently, the statistical study performed by \citet{Krupar:2014aa} on 152 type III bursts at long wavelengths observed by STEREO/SWAVES also showed that the mean type III radio flux increases significantly from 10 MHz to 1 MHz.

\subsection{Interplanetary bursts and $>25$ keV electrons}

As discussed in Section \ref{sec:T3I3T} and summarized in Table \ref{tab:HXRIT3}, the coronal type III bursts in our sample are more often associated with IP type III bursts when they are also associated with detectable 25-50 keV emission. Figure \ref{fig:xray_hist} furthermore shows that for the HXR events in this study that produce more than 250 counts/s in the 25-50 keV channel, there is a much higher chance of detecting an IP burst in connection with the coronal burst.    

The production of high X-ray count rates at 25-50~keV requires a high number of injected electrons above 25 keV.  Increasing the number of injected electrons above 25~keV can be achieved by increasing the number density of electrons at all energies or by hardening the energy distribution (smaller spectral index) of the accelerated electron beam.  A larger number of high-energy electrons increases the Langmuir wave energy obtained from the unstable electron beam.  From quasilinear theory \citep{Vedenov:1963aa,Drummond:1964aa}, the growth rate of Langmuir waves is 
\begin{equation}\label{eqn:growthrate}
\gamma_{ql}=\frac{\pi\omega_{pe}v^2}{n_e}\frac{\partial f}{\partial{v}}.
\end{equation} 
for an electron distribution $f(v)$.  An increase in the density of beam electrons or an increase in the velocity of the electrons increases $\gamma_{ql}$.  Increasing Langmuir wave energy density likely increases type III radio flux \citep[e.g.][]{Melrose:1986aa} and increases the probability of generating a detectable type III burst.  Other factors can inhibit interplanetary type III production (e.g. magnetic connectivity), so we do not expect all beams with a large electron flux above 25~keV to produce interplanetary type III bursts.

The increase in type III flux when the number of electrons above 25~keV is increased was demonstrated numerically by \citet{Li:2008ac,Li:2009aa,Li:2011ab} using a hot, propagating Maxwellian.  They showed that increasing the temperature of the initial Maxwellian beam increases the type III radio flux and increases the bandwidth; the burst starts at higher frequencies and stops at lower frequencies.  The simulations were restricted to frequencies above 150~MHz and the type III flux peaked above 200 MHz, which is inconsistent to the general trend of increasing flux with decreasing frequency reported in this and other studies.  A further study by \citet{Li:2013aa} demonstrated via an initial power-law electron beam that decreasing the power-law spectral index increased the fundamental radio flux emitted at high and low radio frequencies.  Both sets of simulations can explain why a coronal burst produced by an electron beam with a smaller (harder) power-law spectral index, i.e. associated with a larger 25-50 keV count rate, would have a higher probability to be associated with an interplanetary type III burst than if the electron beam
had a larger (softer) initial power-law spectral index.

Electron beams can also be diluted as the cross-section of the guiding magnetic flux tube increases with height.  A high density of electrons above 25 keV is more likely to still produce detectable type III radio flux at altitudes related to 14 MHz, even in the case of a diverging magnetic flux tube.  The effect of flux tube radial expansion has been shown recently by \citet{Reid:2015aa} on type III stopping frequencies. The numerical simulations of beam electrons, and their corresponding resonant interaction with Langmuir waves in diverging magnetic flux tubes, are used to compute the type III stopping frequency; this is defined as the frequency in which the beam is no longer able to produce a sufficient level of Langmuir wave energy density as compared to the thermal level.  Denser electron beams and harder electron beams (low spectral index) are more likely to produce significant Langmuir wave energy densities further away from the injections site than sparse or softer electron beams, and therefore produce type III bursts with lower stopping frequencies. This result provides further understanding as to why coronal bursts associated with stronger HXR bursts are more likely associated with IP type III bursts.

\subsection{Interplanetary bursts and magnetic connectivity}

We found the detection of a strong radio flux at frequencies around 170 MHz (Figure \ref{fig:radio_hist}) to be a very good indication that the type III burst will become an interplanetary type III burst at lower frequencies.  Previously it has not been clear whether the absence of radio emission below 14 MHz is from a weak beam or unfavourable magnetic connectivity.  When electron beams contain enough density to produce high flux type IIIs around 170 MHz we often observe them below 14 MHz.  We deduce that magnetic connectivity plays less of an effect on the transport of electrons from the high corona into interplanetary space.  A strong radio flux at frequencies around 170 MHz indicates the electrons \emph{do} have access to the high corona and subsequently are more likely to access the interplanetary medium.

We did not find the same trend that strong radio flux is a good indication of interplanetary type III bursts at frequencies at and above 237~MHz.  The magnetic connectivity of bursts with a large radio flux at high frequencies, which are produced low in the solar atmosphere, is not necessarily favourable for the electron beam exciter to escape into the upper corona and interplanetary space  (see e.g.\ the example in \citet{Vilmer:2003aa}), even if the radio flux is high.  The access of particles to open field lines and then the possible association between coronal and interplanetary type III bursts may evolve during flares due to, for example, processes of interchange reconnection in which newly emerging flux tubes can reconnect with previously open field lines \citep[see e.g.][]{Masson:2012aa,Krucker:2011aa}.

The statistical results presented in this paper were based on radio spectra and flux time profiles but did not include spatially resolved observations.  This latter aspect will be considered in a following study, which will examine in detail the combination of HXR images provided by RHESSI with the multi-frequency images of the radio bursts produced in the decimeter/meter wavelengths by the Nan\c{c}ay Radioheliograph. Tracing the magnetic connectivity between the solar surface, the corona and the interplanetary medium will be one of the key questions of the Solar Orbiter mission. As shown in the present paper, X-ray and radio emissions from energetic electron beams can be used to trace the electron acceleration and propagation sites from the solar surface to the interplanetary medium. The combination of ground-based radio spectrographs and imagers with the radio, X-ray, and in-situ electron measurements aboard Solar Orbiter will undoubtedly largely contribute in the next decade to a better understanding of the magnetic connectivity between the Sun and the interplanetary medium, and on the release and distribution in space and time of the energetic particles from the Sun.

\begin{acknowledgements}
Hamish Reid acknowledges the financial support from a SUPA Advanced Fellowship and from the STFC consolidated grant ST/L000741/1.  Nicole Vilmer acknowledges support from the Centre National d'Etudes Spatiales (CNES) and from the French programme on Solar-Terrestrial Physics (PNST) of INSU/CNRS for the participation to the RHESSI project.  The European Commission is acknowledged for funding from the HESPE Network (FP7-SPACE-2010-263086).  Financial support by the Royal Society grant (RG130642) is gratefully acknowledged.  Support from a Marie Curie International Research Staff Exchange Scheme RadioSun PEOPLE-2011-IRSES-295272 RadioSun project is greatly appreciated.  Collaborative work was supported by a British council Franco-British alliance grant and funding from the Paris Observatory.  The NRH is funded by the French Ministry of Education and the R\'{e}gion Centre.  The Institute of Astronomy, ETH Zurich and FHNW Windisch, Switzerland is acknowledged for funding Phoenix spectrometers.  The RHESSI team, the WIND/WAVES team and the DAM team are acknowledged for providing data access and analysis software.  
\end{acknowledgements}

\bibliographystyle{aa}
\bibliography{/Users/hamish/Documents/Papers/ubib}

\end{document}